\documentclass[useAMS]{mn2e}

\newcounter{qub}
\newcommand{\qq}{\addtocounter{qub}{1}\arabic{qub}}

\usepackage{graphicx}
\usepackage{txfonts}
\usepackage{rotating}
\def\apgt{\ {\raise-.5ex\hbox{$\buildrel>\over\sim$}}\ }
\def\aplt{\ {\raise-.5ex\hbox{$\buildrel<\over\sim$}}\ }

\title[Revealing evolved massive stars with {\it Spitzer}]{Revealing evolved massive
stars with {\it Spitzer}\footnotemark[0]\thanks{Partially based on observations collected
at the German-Spanish Astronomical Center,
Calar Alto, jointly operated by the Max-Planck-Institut f\"ur Astronomie
Heidelberg and the  Instituto de Astrof\'isica de Andaluc\'ia (CSIC).}}
\author[V.V.Gvaramadze, A.Y.Kniazev and S.Fabrika]
       {V. V.~Gvaramadze$^{1,2}$\thanks{E-mail: vgvaram@mx.iki.rssi.ru },
       A. Y.~Kniazev$^{3,4}$\thanks{E-mail: akniazev@saao.ac.za} and
       S.~Fabrika$^{5}$\thanks{E-mail: fabrika@sao.ru}\\
       $^{1}$Sternberg Astronomical Institute, Moscow State University,
       Universitetskij Pr. 13, Moscow 119992, Russia\\
       $^{2}$Isaac Newton Institute of Chile, Moscow Branch, Universitetskij Pr. 13, Moscow 119992,
       Russia\\
       $^{3}$South African Astronomical Observatory, PO Box 9, 7935 Observatory, Cape Town,
       South Africa \\
       $^{4}$Southern African Large Telescope Foundation, PO Box 9, 7935 Observatory, Cape Town,
       South Africa \\
       $^{5}$Special Astrophysical Observatory, Nizhnij Arkhyz, 369167, Russia
       }
\begin{document}

\date{Accepted 2010 February 8. Received 2010 January 4; in original form 2009 August 1}


\maketitle

\begin{abstract}
Massive evolved stars loss a large fraction of their mass via
copious stellar wind or instant outbursts and during certain
evolutionary phases they can be identified via the presence of their
circumstellar nebulae. In this paper, we present the results of
search for compact nebulae (reminiscent of circumstellar nebulae
around evolved massive stars) using archival 24 $\mu$m data obtained
with the Multiband Imaging Photometer for {\it Spitzer}. We
discovered 115 nebulae, most of which bear a striking resemblance to
the circumstellar nebulae associated with Luminous Blue Variables
(LBVs) and late WN-type (WNL) Wolf-Rayet (WR) stars in the Milky Way
and the Large Magellanic Cloud (LMC). We interpret this similarity
as an indication that the central stars of detected nebulae are
either LBVs or related evolved massive stars. Our interpretation is
supported by follow-up spectroscopy of two dozens of these central
stars, most of which turns out to be either candidate LBVs (cLBVs),
blue supergiants or WNL stars. We expect that the forthcoming
spectroscopy of the remaining objects from our list, accompanied by
the spectrophotometric monitoring of the already discovered cLBVs,
will further increase the known population of Galactic LBVs, which
in turn would have profound consequences for better understanding
the LBV phenomenon and its role in the transition between hydrogen
burning O stars and helium burning WR stars. We also report the
detection of an arc-like structure attached to the cLBV HD\,326823
and an arc associated with the LBV R99 (HD\,269445) in the LMC.
\end{abstract}


\section{Introduction}
%

Massive stars are sources of strong stellar wind whose interaction
with the ambient medium results in formation of circumstellar and
interstellar nebulae (e.g. Johnson \& Hogg 1965; Avedisova 1972;
Weaver et al. 1977). The most well-defined (i.e. compact and
possessing a high degree of symmetry) nebulae are formed during a
brief post-main-sequence phase when (depending on the initial mass)
a massive star becomes a Wolf-Rayet (WR) or an Luminous Blue
Variable (LBV) one (e.g. Vanbeveren, De Loore \& Van Rensbergen
1998; Meynet \& Maeder 2003). Most of WR ring-like nebulae are
formed via the interaction between the fast WR wind and the slow
material shed during the preceding red supergiant phase (e.g.
Chevalier \& Imamura 1983; Marston 1995; Chen, Wang \& Qu 1995;
Brighenti \& D'Ercole 1995, 1997; Garcia-Segura, Langer \& Mac Low
1996a; Smith 1996). These {\it circumstellar} nebulae are observed
exclusively around WR stars of nitrogen sequence and the majority of
them are associated with late WN-type (WNL) stars (e.g. Lozinskaya
\& Tutukov 1981; Esteban et al. 1993), i.e. with young WR stars
whose wind still interacts with the dense circumstellar material
(Gvaramadze et al. 2009a). The nebulae associated with LBVs show a
wide range of morphologies, from ring-like shells to bipolar lobes
and triple-ring systems (e.g. Nota et al. 1995; Weis 2001; Smith
2007). It is believed that the LBV nebulae are formed in sporadic,
violent mass-loss events when the evolved massive stars lose a
significant fraction of their mass (e.g. Smith \& Owocki 2006),
although their formation via continuous outflows is possible too
(e.g. Robberto et al. 1993; Nota et al. 1995; Garcia-Segura, Langer
\& Mac Low 1996b; Voors et al. 2000).

The circumstellar nebulae associated with LBV and WNL stars are
known to be sources of mid- and far-infrared (IR) emission (e.g.
Mathis et al. 1992; Hutsemekers 1997; Trams, Voors \& Waters 1998;
Marston et al. 1999; Egan et al. 2002; Clark et al. 2003) and
therefore can be detected in the archival data of the {\it Spitzer
Space Telescope}. Detection of nebulae reminiscent of those
associated with LBV, WNL and related stars would serve as an
indication that their central stars are evolved massive ones, while
spectroscopic follow-ups of objects selected in this way would allow
us to verify their nature (Clark et al. 2003; Gvaramadze et al.
2009a,b,c).

Until recently, the main channel for detection of new LBV and WR
stars in the Milky Way and nearby galaxies was through the
narrow-band optical surveys accompanied by spectroscopic follow-ups
(e.g. Corral 1996; King, Walterbos \& Braun 1998; Shara et al. 1999;
Sholukhova, Fabrika \& Vlasyuk 1999; Massey et al. 2007). In the
Galaxy, however, the optical surveys can reveal only relatively
nearby objects, whose emission does not significantly suffer from
the interstellar extinction in the Galactic plane. For more distant
objects, the breakthrough results were achieved by means of
narrow-band IR surveys, which allow to select candidate evolved
massive stars via their near- and mid-IR colours almost throughout
the whole Galactic disk. The recent efforts in this direction by
Homeier et al. (2003), Hadfield et al. (2007), Shara et al. (2009)
and Mauerhan, Van Dyk \& Morris (2009) lead to the discovery of 72
new WR stars, i.e. resulted in one third increase of the known
Galactic population of these objects. Many more WR and LBV stars,
however, remain to be discovered (Smith, Crowther \& Prinja 1994;
Shara et al. 1999; van der Hucht 2001; van Genderen 2001).

In this paper we present a list of 115 compact nebulae
(Section\,\ref{sec:list}) detected in the archival data of the {\it
Spitzer Space Telescope}, whose morphology is similar to that of
nebulae associated with known LBVs and related objects
(Section\,\ref{sec:search}; cf. Carey et al. 2009; Wachter et al.
2009). We believe that the majority of these nebulae is produced by
evolved massive stars. Our belief is supported by spectroscopic
follow-ups of two dozens of central stars (CSs) of the nebulae from
our list, which showed that most of them are either candidate LBVs
(cLBVs), blue supergiant or WNL stars (Section\,\ref{sec:spectr};
Gvaramadze et al., in preparation; Kniazev et al., in preparation).

\section{Search for infrared nebulae}
\label{sec:search}

During the search for bow shocks around OB stars ejected from young
star clusters [for motivation of this search see Gvaramadze \&
Bomans (2008)] using the archival data of the {\it Spitzer Space
Telescope} (Werner et al. 2004), obtained with the Multiband Imaging
Photometer for {\it Spitzer} (MIPS; Rieke et al. 2004) and the
Infrared Array Camera (IRAC; Fazio et al. 2004), we serendipitously
discovered several compact ($\sim 1\arcmin -2\arcmin$ in size)
nebulae (see Fig.\,\ref{fig:new}), whose morphology is extremely
similar to that of nebulae associated with known (c)LBVs and WNL
stars in the Milky Way and the Large Magellanic Cloud (LMC) (e.g.
Marston et al. 1999; Voors et al. 2000; Egan et al. 2002; Clark et
al. 2003; Weis 2003). Using the SIMBAD
database\footnote{http://simbad.u-strasbg.fr/simbad/}, we found that
the CS associated with one of the nebulae (MN114; see Table\,1 of
Section\,\ref{sec:list}) was identified by Dolidze (1971) as a
possible WR star due to the presence in its spectrum of an emission
band around $\lambda \, 6750$ \AA. Our spectroscopic follow-up
(carried out with the Russian 6-m telescope in 2008
October-November) confirmed the WR nature of this star and showed
that it belongs to the WN8-9h subtype (Gvaramadze et al. 2009a; see
also Section\,\ref{sec:spectr}). On this basis, we interpreted the
discovered nebulae as the products of post-main-sequence evolution
of massive stars (cf. Carey et al. 2009).

\begin{figure}
\caption{{\it Spitzer} MIPS $24 \, \mu$m images of representative
newly discovered nebulae: a) MN1; b) MN7; c) MN21; d) MN23; e) MN27;
f) MN30; g) MN43; h) MN46; i) MN48; j) MN55; k) MN59; l) MN79; m) MN82;
n) MN96; o) MN98; p) MN108; q) MN112; r) MN113. Here
and hereafter the coordinates are in units of RA(J2000) and Dec.(J2000)
on the horizontal and vertical scales, respectively.
}
\label{fig:new}
\end{figure}

To substantiate this interpretation, we searched for IR
circumstellar nebulae around the known (c)LBVs and WN stars in the
Milky Way and the LMC [see, respectively, Clark, Larionov \&
Arkharov (2005) and Massey, Waterhouse \& DeGioia-Eastwood (2000)
for a recent census of Galactic and LMC (c)LBVs, and van der Hucht
(2001, 2006) and Breysacher, Azzopardi \& Testor (1999) for
catalogue of WR stars] using the MIPS $24\,\mu$m data. The detected
nebulae are shown on Fig.\,\ref{fig:known}.
\begin{figure}
\caption{{\it Spitzer} MIPS $24 \, \mu$m [panels a)~--j) and
p)~--r)] and IRAC $8.0\,\mu$m [panels k)~--o)] images of nebulae
associated with Galactic [panels a)~--o)] and LMC [panels p)~--r)]
(c)LBVs and WN stars: a) cLBV Hen\,3-519 or WR\,31a (WN11h), b) WR\,75ab
(WN7h), c) cLBV HD\,326823, d) cLBV Wray\,17-96, e) cLBV HD\,316285,
f) LBV GAL\,024.73+00.69; g) cLBV GAL\,026.47+00.02, h) WR\,124 (WN8h),
i) WR\,136 (WN6), j) cLBV GAL\,079.29+00.46, k) LBV AG\,Car or WR\,31b
(WN11h), l) cLBV Sher\,25; m) cLBV Pistol; n) cLBV HD\,168625; o) cLBV GAL\,025.5+00.2;
p) Brey\,2 (WN4), q) Brey\,13 (WN8), r) LBV R\,99 (HD\,269445).
} \label{fig:known}
\end{figure}
All but two these nebulae were previously detected either in the
optical, radio or IR. To our knowledge, an arc-like nebula attached
to the cLBV HD\,326823 (Fig.\,\ref{fig:known}c) and a bow shock-like
structure to the east of the LMC LBV R99 (HD\,269445)
(Fig.\,\ref{fig:known}r) are detected for the first time [cf.,
respectively, Clark et al. (2005) and Weis (2003)]. We note the
existence of a halo around the already known IR shell (Trams et al.
1998; Egan et al. 2002) surrounding the cLBV GAL\,079.29+00.46 (see
Fig.\,\ref{fig:known}j). [This halo was independently detected by
Kraemer et al. (2009)]. We also note the clear bipolar appearance of
the nebula (known as M\,1-67) associated with the WN8h star WR\,124
(see Fig.\,\ref{fig:known}h). The existence of a bipolar outflow in
M\,1-67 was claimed by Sirianni et al. (1998) on the basis of
long-slit spectroscopy of this nebula and subsequently questioned by
Grosdidier et al. (2001). Detection of bipolar protrusions in the
shell of M\,1-67 allows us to confirm the claim by Sirianni et al.
(1998). The MIPS $24 \, \mu$m images of nebulae associated with the
LBV AG\,Car and four cLBVs, Pistol Star, Sher\,25, HD\,168625 and
GAL\,025.5+00.2, are saturated and we give their IRAC $8.0\,\mu$m
images. The {\it Spitzer} images of HD\,168625, HD\,316285 and
GAL\,025.5+00.2 were earlier presented by, respectively, Smith
(2007), Morris (2008) and Phillips \& Ramos-Larios (2008a).

\begin{figure}
\begin{center}
\includegraphics[width=1.0\columnwidth,angle=0]{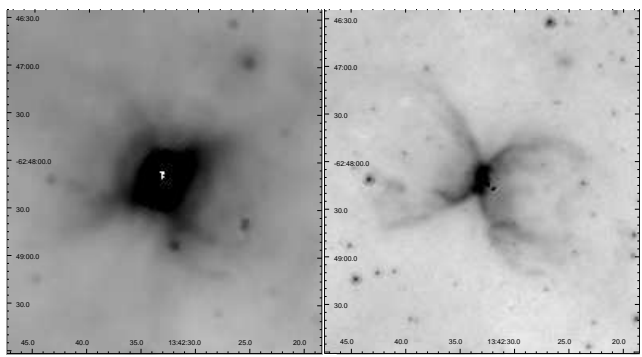}
\end{center}
\caption{{\it Spitzer} MIPS $24 \, \mu$m (left) and IRAC $8 \, \mu$m images of MN13.}
\label{fig:J134233(24+8)}
\end{figure}

Comparison of Fig.\,\ref{fig:known} with Fig.\,\ref{fig:new} shows a
striking similarity between the majority of known and newly
discovered nebulae. For example, we draw attention to the nebula
MN112 (Fig.\,\ref{fig:new}q) associated with the newly detected cLBV
(see Section\,\ref{sec:spectr} for more details), whose shell-halo
morphology is reminiscent of that of the nebula of cLBV
GAL\,079.29+00.46 (Fig.\,\ref{fig:known}j). Other striking examples
are the bipolar nebulae MN79 (Fig.\,\ref{fig:new}l) and MN13
(Fig.\,\ref{fig:J134233(24+8)}) which show a strong resemblance to
the nebulae around, respectively, the WNL star WR\,124
(Fig.\,\ref{fig:known}h) and the cLBV HD\,168625
(Fig.\,\ref{fig:known}n). This similarity provides further evidence
that the discovered nebulae could be products of post-main-sequence
evolution of massive stars and motivated us for a systematic search
for similar nebulae in the {\it Spitzer Space Telescope} archival
data. The goal of the search is to identify new LBVs and related
stars via detection of compact IR nebulae and spectroscopic
follow-up of their CSs.
\begin{figure}
\begin{center}
\includegraphics[width=1.0\columnwidth,angle=0]{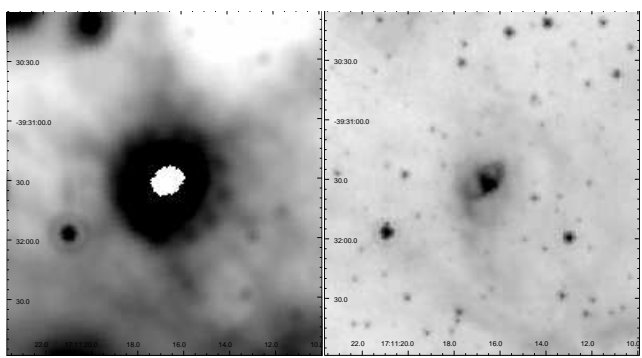}
\end{center}
\caption{{\it Spitzer} MIPS $24 \, \mu$m (left) and IRAC $5.8 \, \mu$m images of MN56.}
\label{fig:J171116(24+5.8)}
\end{figure}

Since we know from experience that most of newly detected nebulae
are visible only at $24\,\mu$m (cf. Carey et al. 2009), we used for
the search the MIPS $24\,\mu$m data obtained within the framework of
the 24 and 70 Micron Survey of the Inner Galactic Disk with MIPS
(MIPSGAL; Carey et al. 2009) and the Cygnus-X Spitzer Legacy
Survey\footnote{http://www.cfa.harvard.edu/cygnusX/index.html} (Hora
et al. 2009). The first survey mapped 278 square degrees of the
inner Galactic plane: $-65\degr < l < -10\degr$ and $10\degr < l
<65\degr$ for $|b| < 1\degr$, while the second one covers 24 square
degrees in Cygnus\,X -- one of the most massive star forming
complexes in the Milky Way (e.g. Reipurth \& Schneider 2008). The
MIPS imaging data were retrieved from the {\it Spitzer} archive
using the Leopard software. The visual inspection of the data
revealed a multitude of compact nebulae possessing one or another
type of symmetry, ranging from pure circular to bilateral or bipolar
(cf. Billot et al. 2009). From these nebulae we selected the ones
with detectable CSs, which constitute our sample for spectroscopic
follow-up.

To identify the CSs we used the Two Micron All Sky Survey (2MASS;
Skrutskie et al. 2006), which provides images in the three near-IR
$J(1.25 \, \mu$m), $H(1.65 \, \mu$m), and $K_s(2.16 \, \mu$m)
bandpasses with the limiting magnitudes of 16.4, 15.5 and 14.8,
respectively. Most of the identified CSs are highly-reddened ($J-K_s
\ga 1$ mag) objects with no optical counterparts in the Digitized
Sky Survey II (DSS-II; McLean et al. 2000), which covers 98 per cent
of the celestial sphere in the $R$ band and provides images to the
limiting magnitude of $\sim 21$. Six CSs are marginally detected or
absent on 2MASS images, while visible at IRAC (3.6, 4.5, 5.8 and
$8.0 \, \mu$m) images obtained within the framework of the Galactic
Legacy Infrared Mid-Plane Survey Extraordinaire (GLIMPSE; Benjamin
et al. 2003).

\begin{figure}
\begin{center}
\includegraphics[width=1.0\columnwidth,angle=0]{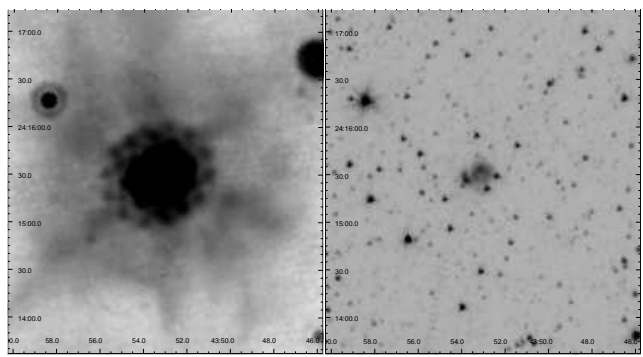}
\end{center}
\caption{{\it Spitzer} MIPS $24 \, \mu$m (left) and IRAC $3.6 \, \mu$m images of MN111.}
\label{fig:J194353(24+3.6)}
\end{figure}

The selected nebulae are listed in Table\,1 of
Section\,\ref{sec:list}. We caution that this list should not be
considered complete. Detection of weak and small-scale nebulae is
hampered by the complex environment of star-forming regions (see,
e.g., Fig.\,\ref{fig:new}f), while identification of bright objects
might be precluded due to the saturation of MIPS $24\, \mu$m images.
In the latter case, inspection of IRAC images of the saturated MIPS
sources could be useful. Several interesting objects revealed in
this way are MN2, MN4, MN56, MN83 and MN111 (see
Fig.\,\ref{fig:J171116(24+5.8)} and
Fig.\,\ref{fig:J194353(24+3.6)}).
\begin{figure}
\begin{center}
\includegraphics[width=1.0\columnwidth,angle=0]{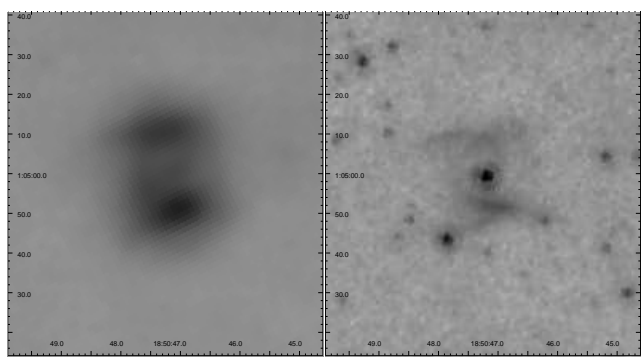}
\end{center}
\caption{{\it Spitzer} MIPS $24 \, \mu$m (left) and IRAC $8 \, \mu$m images of MN94.}
\label{fig:J185047(24+8)}
\end{figure}

The majority of rejected (sourceless) nebulae are compact ($\la 0.5$
arcmin in diameter) circular objects. Some of them have limb
brightened or bilateral morphology, while others exhibit more
complex brightness distribution. A first example of sourceless
nebulae discovered with the MIPS was reported by Morris et al.
(2006), who found a bilateral shell in the Galactic First Look
Survey with the {\it Spitzer Space Telescope}. This nebula (like
most of the nebulae detected in the {\it Spitzer} data) is visible
only at $24\,\mu$m. The IR spectroscopy of the nebula showed that
its emission arises almost solely from the [O\,{\sc iv}] $25.89 \,
\mu$m emission line, which explains the unusual colour of the
nebula. The most likely interpretation of the shell is that it is a
distant young supernova remnant (Morris et al. 2006). It is
possible, therefore, that some of the (sourceless) nebulae we
rejected are also the distant supernova remnants. On the other hand,
some of them could be very distant circumstellar nebulae, whose
highly-reddened CSs are beyond the detection limits of the current
IR surveys.

\begin{figure}
\begin{center}
\includegraphics[width=1.0\columnwidth,angle=0]{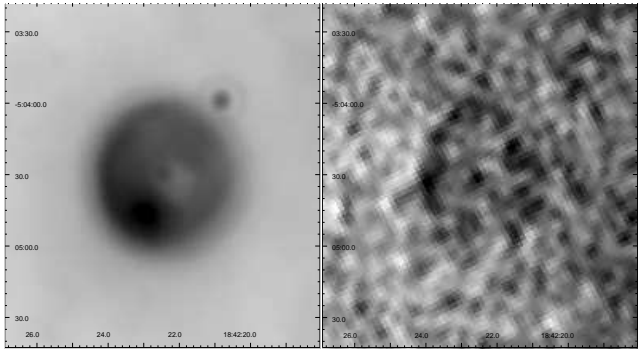}
\end{center}
\caption{{\it Spitzer} MIPS $24 \, \mu$m (left) and MAGPIS 20 cm images of MN87.}
\label{fig:J184222(24+radio)}
\end{figure}

\begin{figure}
\begin{center}
\includegraphics[width=1.0\columnwidth,angle=0]{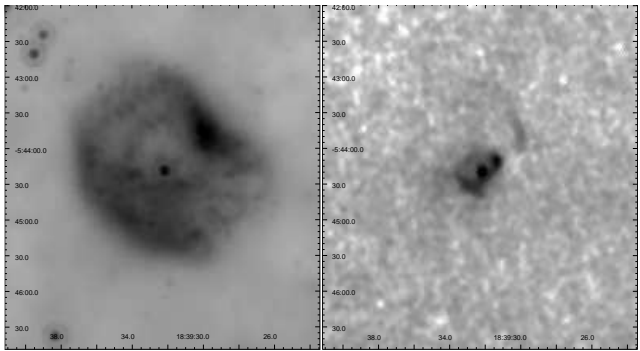}
\end{center}
\caption{{\it Spitzer} MIPS $24 \, \mu$m (left) and MAGPIS 20 cm images of the cLBV GAL\,026.47+00.02.}
\label{fig:GAL026.47+00.02(24+radio)}
\end{figure}

All but two of the nebulae presented in Table\,1 were detected in
the data of the MIPSGAL survey; the high resolution images of these
nebulae are available in the NASA/IPAC Infrared Science
Archive\footnote{http://irsa.ipac.caltech.edu/index.html}. The
majority of the nebulae are visible only at MIPS $24 \, \mu$m images
and few of them have counterparts in the IRAC wavebands
($3.6-8.0\,\mu$m). In some cases, the IRAC images show more
spectacular structures than the MIPS $24\, \mu$m ones (see
Fig.\,\ref{fig:J134233(24+8)} and Fig.\,\ref{fig:J185047(24+8)}).
Two nebulae, MN18 and MN102 (see Table\,1), have optical
counterparts, respectively, in the SuperCOSMOS H-alpha Survey (SHS;
Parker et al. 2005) and the DSS-II. MN18 is associated with an early
B supergiant (see Section\,\ref{sec:spectr}), while MN102 is most
likely a planetary nebula (PN) with the [WC] CS (Gvaramadze et al.,
in preparation). Nebulae located in the region $5\degr <l<48.5\degr,
|b| < 0.8\degr$ are covered by the Multi-Array Galactic Plane
Imaging Survey (MAGPIS; Helfand et al. 2006). Four of them (MN83,
MN86, M87 and MN101) have radio counterparts at 20 cm (e.g.
Fig.\,\ref{fig:J184222(24+radio)}). For the sake of completeness, we
note that MAGPIS covers several known (c)LBVs, of which the cLBV
GAL026.47+00.02 has a very interesting radio counterpart
(Fig.\,\ref{fig:GAL026.47+00.02(24+radio)}), whose appearance
strikingly resembles the tore-like structure of MN56 (see right
panel of Fig.\,\ref{fig:J171116(24+5.8)}).
\begin{table*}
\caption{List of MIPS $24\,\mu$m nebulae.} \label{catalog}
\centering
\renewcommand{\footnoterule}{}  
\begin{tabular}{lccccccccccccc}
\hline
Name & RA (J2000) & Dec. (J2000) & $B$ & $V$ & $R$ & $J$ & $H$ & $K_s$ & Type & Size ($''\times ''$) & Comments \\
\hline
MN\qq & 11 44 18.10 & -62 45 20.2 & 17.37 & 17.04 & 14.85 & 10.00 & 8.72 & 7.78 & E & 60$\times$70 & \\
MN\qq    & 12 00 58.82 & -63 12 59.9 & -- & 13.23 & -- & 13.29 & 12.65 & 12.53 & R,BL & 30 & \\
MN\qq & 12 10 13.37 & -62 50 58.3 & -- & -- & -- & 16.08 & 14.47 & 14.13 & R & 20 & \\
MN\qq & 12 17 10.37 & -62 06 27.1 & -- & -- & -- & 13.94 & 12.58 & 11.98 & BL & 35$\times$45  & \\
MN\qq & 12 29 41.70 & -62 13 08.4 & 16.60 & 16.71 & 17.47 & 14.94 & 13.74 & 12.20 & C & -- & \\
MN\qq & 12 47 17.74 & -63 00 34.3 & -- & -- & -- & 15.62 & 15.07 & 14.73 & R,BL & 25 &  \\
MN\qq & 13 10 04.87 & -63 11 30.1 & 15.20 & 14.38 & 12.83 & 9.54 & 7.95 & 6.38 & E & 10$\times$20 & Wray 16-126 (Be)$^{(1)}$ \\
MN\qq & 13 10 43.84 & -63 17 45.8 & -- & 16.21 & -- & 12.27 & 11.00 & 10.19 & R,BL & 55 &  \\
MN\qq & 13 14 57.05 & -62 23 53.3 & -- & -- & -- & 13.83 & 11.88 & 10.67 & E & 15$\times$25 &  \\
MN\qq & 13 19 33.87 & -62 38 44.8 & 10.98 & 10.36 & 9.94 & 8.25 & 8.01 & 7.84 & R & 40 & CD-61 3738 (B5Iap)$^{(2)}$ \\
MN\qq & 13 27 08.11 & -62 03 19.8 & -- & -- & 17.62 & 12.92 & 11.95 & 11.28 & BP & 30$\times$45 &  \\
MN\qq & 13 27 17.85 & -63 19 28.1 & -- & -- & -- & 14.52 & 14.17 & 13.54 & BP & 30$\times$45&  \\
MN\qq & 13 42 33.08 & -62 48 11.3 & -- & -- & -- & 15.15 & 13.27 & 10.07 & BP & 140 &  \\
MN\qq & 14 31 11.00 & -61 02 02.1 & 14.22 & 13.27 & 12.98 & 11.80 & 11.41 & 11.03 & BP & 60 &  \\
MN\qq & 15 01 35.02 & -58 00 26.1 & -- & -- & -- & 15.39 & 14.04 & 13.30 & R,BL & 40 &  \\
MN\qq & 15 07 56.37 & -58 18 23.8 & 17.47 & 17.34 & 16.58 & 9.28 & 7.82 & 6.96 & R & 64 & \\
MN\qq & 15 13 42.17 & -58 53 18.5 & 18.58 & -- & 17.37 & 13.31 & 12.49 & 12.05 & BL & 20  & \\
MN\qq & 15 16 41.00 & -58 22 26.0 & 14.59 & 12.84 & 11.98 & 9.22 & 8.72 & 8.38 & BP & 100$\times$150 & blue supergiant$^{(3)}$ \\
MN\qq & 15 19 59.90 & -57 24 15.2 & 16.86 & 15.64 & 14.78 & 12.77& 12.35 & 11.99 & R,BL & 40 & \\
MN\qq & 15 31 41.30 & -56 08 53.0 & -- & -- & -- & -- & -- & -- & R & 25 & weak 2MASS $K_s$ source  \\
MN\qq & 15 33 08.10 & -56 12 20.0 & -- & -- & -- & 11.68 & 9.69 & 8.57 & R & 90 & \\
MN\qq & 15 34 12.01 & -55 29 57.4 & -- & -- & -- & 16.19 & 14.66 & 13.74 & R,BL & 25 & \\
MN\qq & 15 35 26.53 & -56 04 12.3 & -- & -- & -- & 13.84 & 12.39 & 11.46 & R & 140 & WN7 star$^{(4)}$ \\
MN\qq & 15 35 53.00 & -55 57 04.6 & -- & -- & -- & 13.43 & 11.86 & 11.25 & R & 80 & \\
MN\qq & 15 38 56.77 & -56 37 22.9 & 17.87 & -- & -- & 14.65 & 14.08 & 13.86 & R,BL & 20 &  \\
MN\qq & 15 45 27.46 & -53 56 02.6 & 19.70 & -- & 18.81 & 15.73 & 14.98 & 14.11 & E & 55$\times$70 & \\
MN\qq & 15 45 59.14 & -53 32 32.5 & -- & -- & -- & 14.58 & 12.22 & 10.87 & R & 100 &  \\
MN\qq & 15 46 51.72 & -53 44 58.8 & -- & -- & -- & 17.78 & 16.53 & 15.02 & R & 70 &  \\
MN\qq & 15 48 36.40 & -54 30 46.9 & -- & -- & -- & 17.83 & 12.24 & 9.48 & R & 80 & \\ 
MN\qq & 15 48 42.07 & -55 07 42.2 & 18.94 & 16.52 & 14.51 & 7.93 & 6.73 & 5.98 & R & 215 &   \\
MN\qq & 15 49 05.43 & -53 44 12.8 & -- & -- & -- & 13.40 & 9.74 & 7.79 & R & 35 &  \\
MN\qq & 15 49 25.97 & -55 13 41.6 & 20.40 & -- & 19.21 & 15.34 & 13.55 & 12.68 & R & 50 &  \\
MN\qq & 15 52 54.77 & -52 59 47.0 & -- & -- & -- & 15.81 & 14.10 & 12.82 & BL & 20 &  \\
MN\qq & 15 58 11.48 & -52 55 50.4 & -- & -- & -- & -- & -- & -- & R & 90 & [3.6]=11.80 mag \\
MN\qq & 15 58 13.78 & -52 57 51.4 & -- & -- & -- & 13.77 & 11.37 & 9.98 & R & 120 &  \\
MN\qq & 15 58 45.84 & -54 11 48.9 & 18.53 & -- & 16.67 & 13.89 & 13.12 & 12.66 & R & 40 & \\
MN\qq & 15 59 18.88 & -54 07 37.5 & -- & -- & -- & -- & -- & -- & R & 25 & [3.6]=14.40 mag \\
MN\qq & 16 05 52.99 & -52 50 37.8 & -- & -- & -- & 14.66 & 13.89 & 13.28 & R & 75 & \\
MN\qq & 16 10 26.55 & -51 21 25.3 & -- & -- & -- & 17.65 & 15.36 & 11.26 & R & 35 &  \\
MN\qq & 16 11 32.21 & -51 29 06.4 & 18.25 & 15.52 & 14.82 & 10.62 & 10.08 & 9.80 & R & 45 &  \\
MN\qq & 16 26 34.28 & -50 21 01.9 & -- & -- & 16.26 & 9.93 & 8.55 & 7.74 & R & 50 & PN(?)$^{(5)}$ \\
MN\qq & 16 31 37.82 & -48 14 55.3 & -- & -- & -- & 16.57 & 12.60 & 10.28 & E & 40$\times$45 & \\
MN\qq & 16 32 14.11 & -47 50 40.1 & -- & -- & -- & 14.20 & 12.50 & 11.79 & R & 135 & WN5$^{(6)}$ \\
MN\qq & 16 32 39.95 & -49 42 13.8 & 16.55 & 16.39 & 14.43 & 8.41 & 7.46 & 6.81 & R & 50 & EM$^*$ VRMF\,55 (emission line star)$^{(7)}$ \\
MN\qq & 16 36 42.78 & -46 56 20.7 & -- & -- & -- & 11.71 & 7.89 &  5.82 & R & 50 & \\
MN\qq & 16 43 16.37 & -46 00 42.4 & 17.12 & 15.66 & 12.92 & 6.26 & 5.08 & 4.21 & E & 120$\times$140 & Hen 2-179; cLBV$^{(8)}$ \\
MN\qq & 16 46 17.35 & -45 08 47.9 & -- & -- & -- & 15.38 & 11.87 & 9.85 & E & 30$\times$45 & \\
MN\qq & 16 49 37.70 & -45 35 59.2 & 17.64 & 15.83 & 13.82 & 7.24 & 6.09 & 5.42 & R & 200 &  \\
MN\qq & 17 07 23.16 & -40 07 41.0 & -- & -- & -- & 12.07 & 9.48 & 7.70 & BP & 40$\times$65 &  \\
MN\qq & 17 07 23.34 & -39 56 50.5 & 19.30 & 16.70 & 13.48 & 7.01 & 5.33 & 4.63 & E & 60$\times$90 &  \\
MN\qq & 17 08 29.13 & -39 25 07.7 & -- & -- & -- & 10.49 & 8.22 & 7.11 & E & 90$\times$100 &  \\
MN\qq & 17 09 22.37 & -40 08 01.2 & -- & -- & -- & 15.44 & 14.10 & 13.04 & R & 20 &  \\
MN\qq & 17 09 24.78 & -40 08 45.6 & -- & -- & -- & 12.98 & 10.84 & 8.53 & R & 50 &  \\
MN\qq & 17 09 33.99 & -40 09 10.1 & -- & -- & -- & 11.70 & 10.31 & 9.75 & E & 25$\times$40&  \\
MN\qq & 17 11 01.00 & -39 45 18.7 & 20.94 & -- & 17.59 & 9.73 & 8.04 & 6.96 & R & 65 &  \\
MN\qq & 17 11 16.69 & -39 31 31.3 & -- & -- & -- & 10.47 & 8.77 & 7.83 & C & -- &  \\
MN\qq & 17 20 08.04 & -36 13 23.9 & 21.22 & -- & 17.51 & 11.78 & 10.66 & 10.06 & E & 45$\times$70 &  \\
MN\qq & 17 37 47.54 & -31 37 33.4 & -- & -- & -- & 11.77 & 9.25 & 7.79 & R & 75 &  \\
MN\qq & 17 39 18.99 & -31 24 24.3 & 17.71 & 15.61 & 13.87 & 8.44 & 7.59 & 7.08 & R,BL & 75 &  \\
MN\qq & 17 42 07.62 & -26 02 13.4 & -- & -- & 18.62 & 13.90 & 12.74 & 12.31 & R & 55 &  \\
MN\qq & 17 42 14.02 & -29 55 36.1 & -- & -- & -- & 13.38 & 10.54 & 8.99 & R & 40 &  \\
MN\qq & 17 42 19.32 & -32 35 32.2 & -- & -- & -- & 13.91 & 12.65 & 12.19 & BP & 45 &  \\
MN\qq & 17 43 09.50 & -33 51 11.0 & -- & -- & -- & -- & -- & -- & R & 25 & [3.6]=13.95 mag \\
MN\qq & 17 43 59.85 & -30 28 38.5 & 17.04 & 13.75 & 11.93 & 5.74 & 4.78 & 4.07 & E & 110$\times$140 & 2MASS J17435981-3028384 (M2)$^{(9)}$\\
\hline
\end{tabular}
\end{table*}
\addtocounter{table}{-1}
\begin{table*}
\caption{(continued)} \label{catalog} \centering
\renewcommand{\footnoterule}{}  
\begin{tabular}{lcccccccccccc}
\hline
Name &  RA(J2000) & Dec.(J2000) & B & V & R & J & H & $K_s$ & Type & Size ($''\times ''$) & Comments \\
\hline
MN\qq & 17 45 40.77 & -27 09 15.0 & -- & -- & 16.98 & 14.59 & 14.49 & 14.01 & R & 25 &  \\
MN\qq & 17 47 05.21 & -27 25 33.5 & -- & -- & 18.52 & 14.84 & 14.98 & 14.38 & BL & 45 &  \\
MN\qq & 17 48 39.21 & -26 23 15.9 & -- & -- & -- & 14.43 & 12.87 & 12.25 & R & 30 &  \\
MN\qq & 18 02 22.34 & -22 38 00.2 & -- & -- & -- & 13.60 & 11.07 & 9.60 & R & 55 &  \\
MN\qq & 18 02 43.91 & -22 37 47.0 & -- & -- & -- & 13.38 & 11.34 & 10.34 & BL & 30$\times$40 &  \\
MN\qq & 18 03 56.68 & -22 56 00.0 & -- & -- & 18.78 & 14.02 & 13.33 & 12.93 & R & 30 &  \\
MN\qq & 18 04 39.66 & -21 41 34.0 & 19.74 & -- & 16.74 & 13.32 & 12.54 & 12.01 & E & 35$\times$45 &  \\
MN\qq & 18 04 44.44 & -21 50 25.7 & -- & -- & -- & 13.33 & 10.58 & 9.20 & E & 15$\times$25 &  \\
MN\qq & 18 05 14.84 & -23 47 09.3 & -- & -- & -- & 14.76 & 13.23 & 12.57 & R & 30 &  \\
MN\qq & 18 06 12.92 & -21 17 45.7 & 11.90 & 11.76 & 11.68 & 11.23 & 10.98 & 10.22 & R & 25 & HD\,313771 (B9)$^{(10)}$ \\
MN\qq & 18 06 36.21 & -19 53 47.3 & -- & -- & -- & 10.83 & 9.10 & 8.01 & R & 30 & PN(?)$^{(11)}$ \\
MN\qq & 18 07 05.17 & -20 15 16.3 & -- & -- & -- & 14.87 & 12.67 & 11.16 & R & 50 &  \\
MN\qq & 18 13 31.20 & -18 56 43.0 & -- & -- & 20.11 & 11.57 & 10.34 & 9.65 & R,BL & 45 &  \\
MN\qq & 18 17 15.54 & -14 53 05.1 & -- & -- & -- & 13.89 & 12.86 & 11.95 & R & 90 &  \\
MN\qq & 18 28 33.41 & -11 46 44.2 & -- & -- & -- & 14.36 & 11.42 & 9.71 & BP & 45$\times$70 &  \\
MN\qq & 18 33 39.55 & -08 07 08.5 & -- & -- & -- & 13.00 & 9.85 & 8.15 & R & 80 &  \\
MN\qq & 18 33 43.47 & -08 23 35.3 & -- & -- & -- & 14.43 & 12.03 & 9.81 & R & 35 &  \\
MN\qq & 18 37 33.49 & -06 46 28.8 & -- & -- & -- & 14.46 & 12.63 & 11.69 & BP & 40$\times$50 &  \\
MN\qq & 18 39 23.01 & -05 53 19.9 & -- & -- & -- & 13.19 & 10.09 & 8.39 & E & 30$\times$45 & LBV(?)$^{(12)}$ \\
MN\qq & 18 41 59.73 & -05 15 39.4 & 18.80 & -- & 15.77 & 7.96 & 6.53 & 5.68 & R & 60 &  \\
MN\qq & 18 42 06.31 & -03 48 22.5 & -- & -- & -- & 11.95 & 10.22 & 9.16 & R & 120 &  \\
MN\qq & 18 42 08.27 & -03 51 02.9 & -- & -- & -- & 11.85 & 10.26 & 9.27 & R & 90 & interacts with MN81\\
MN\qq & 18 42 22.47 & -05 04 30.1 & -- & -- & -- & 13.64 & 11.10 & 8.68 & R & 60 &  \\
MN\qq & 18 42 27.40 & -03 56 34.1 & -- & -- & -- & -- & -- & -- & R & 20 & [3.6]=13.51 mag  \\
MN\qq & 18 44 54.85 & -03 35 38.6 & -- & -- & -- & 14.19 & 12.26 & 10.40 & BP & 15$\times$35 &  \\
MN\qq & 18 45 55.94 & -03 08 29.7 & -- & -- & -- & 15.45 & 11.51 & 9.37 & R & 60 &  \\
MN\qq & 18 49 27.39 & -01 04 20.3 & 18.31 & 16.42 & 14.05 & 10.94 & 10.07 & 9.47 & R & 120 & WR\,121b (WN7h)$^{(13)}$ \\
MN\qq & 18 50 01.97 & -00 56 14.3 & -- & -- & -- & 13.77 & 12.65 & 12.03 & BP & 40$\times$60 &  \\
MN\qq & 18 50 39.80 & 00 04 45.4 & -- & -- & -- & 16.39 & 13.64 & 12.17 & R,BL & 45 &  \\
MN\qq & 18 50 47.24 & 01 04 59.1 & -- & -- & -- & 13.05 & 11.04 & 9.94 & BP & 40 &  \\
MN\qq & 18 51 00.29 & -00 32 31.0 & -- & -- & -- & 14.11 & 12.08 & 11.00 & R & 45 &  \\
MN\qq & 18 51 02.95 & -00 58 24.2 & -- & -- & -- & 11.88 & 10.14 & 9.13 & R & 135 &  \\
MN\qq & 18 53 05.82 & 00 11 35.8 & -- & -- & -- & 14.43 & 12.54 & 11.64 & E & 35$\times$40 &  \\
MN\qq & 19 01 16.69 & 03 55 10.8 & -- & -- & -- & 15.22 & 11.66 & 9.69 & R & 65 &  \\
MN\qq & 19 03 29.45 & 04 07 20.2 & -- & -- & -- & 16.11 & 14.57 & 13.96 & R & 20 &  \\
MN\qq & 19 04 21.07 & 06 00 01.2 & 20.02 & -- & 17.06 & 14.78 & 12.96 & 11.57 & R & 95 &  \\
MN\qq & 19 06 24.54 & 08 22 01.6 & -- & -- & 16.78 & 9.66 & 7.92 & 6.85 & R & 45 &  \\
MN\qq & 19 06 33.65 & 09 07 21.1 & -- & -- & 17.40 & 15.59 & 14.86 & 13.88 & R,BL & 45 & [WC]$^{14}$ \\
MN\qq & 19 07 57.95 & 07 55 21.5 & -- & -- & -- & -- & -- & -- & R & 25 & [3.6]=13.89 mag \\
MN\qq & 19 08 46.19 & 08 43 29.0 & -- & -- & -- & 16.52 & 15.33 & 14.44 & R & 15$\times$20 &  \\
MN\qq & 19 10 04.27 & 10 43 29.2 & -- & -- & -- & 15.21 & 14.68 & 13.75 & R,BL & 40 &  \\
MN\qq & 19 13 32.75 & 08 27 03.1 & -- & -- & 17.21 & 12.38 & 10.98 & 10.45 & R & 70 &  \\
MN\qq & 19 24 03.34 & 13 39 49.4 & 18.96 & 16.78 & 15.29 & 10.16 & 9.21 & 8.60 & R & 40 &  \\
MN\qq & 19 26 58.96 & 18 46 44.0 & 17.69 & 15.52 & 15.23 & 11.66 & 11.11 & 10.79 & BP & 50 & blue supergiant$^{(15)}$ \\
MN\qq & 19 28 14.58 & 17 16 23.1 & -- & -- & -- & 14.44 & 12.19 & 11.06 & E & $15\times20$ & blue supergiant (?)$^{(16)}$ \\
MN\qq & 19 30 33.49 & 18 35 56.7 & 20.50 & 17.97 & 17.29 & 11.96 & 11.04 & 10.55 & BL & 20$\times$25 &  \\
MN\qq & 19 43 53.19 & 24 15 31.3 & 17.59  & 16.64 & 15.60 & 13.91 & 13.12 & 12.20 & C & -- &  \\
MN\qq & 19 44 37.60 & 24 19 05.9 & 16.42 & 14.64 & 13.68 & 8.86 & 8.02 & 7.42 & R & 100 & cLBV$^{(17)}$ \\
MN\qq & 19 44 42.96 & 23 11 33.9 & 15.49 & 14.20 & 13.10 & 9.92 & 9.38 & 9.10 & R & 60 &  \\
MN\qq & 20 17 08.12 & 41 07 27.0 & 16.54 & 15.18 & 14.53 & 10.15 & 9.27 & 8.65 & R & 135 & WR 138a (WN8-9h)$^{(18)}$ \\
MN\qq & 20 35 16.03 & 42 20 16.4 & -- & 17.20 & 17.76 & 11.61 & 9.95 & 8.88 & BP & 75$\times$100 & \\
\hline
\multicolumn{12}{p{18cm}}{%
(1) SIMBAD; (2) SIMBAD; (3) this paper; (4) Mauerhan et al. (2009); (5) Su\'{a}rez et al. (2006); (6) Shara et al. (2009); (7) SIMBAD;
(8) this paper; (9) SIMBAD; (10) SIMBAD; (11) Phillips \& Ramos-Larios (2008b); (12) Davies et al. (2007); (13) Gvaramadze et al.
(2009b); (14) Gvaramadze et al., in preparation; (15) this paper; (16) Phillips \& Ramos-Larios (2008c); (17) Gvaramadze
et al. (2009c); (18) Gvaramadze et al. (2009a). }
\end{tabular}
\end{table*}

\section{List of infrared nebulae}
\label{sec:list}

Table\,1 gives the list of nebulae with CSs discovered in the
archival {\it Spitzer} MIPS $24\,\mu$m data. We initialize nebulae
with MN (i.e. the MIPS nebulae) and assign a sequence number to each
object in column 1. Columns 2 and 3 give the equatorial coordinates
of CSs associated with the nebulae (taken from the 2MASS catalogue;
Skrutskie et al. 2006). Columns 4-6 and 7-9 give the B,V,R and
J,H,K$_{\rm S}$ magnitudes of the CSs, taken, respectively, from the
NOMAD (Zacharias et al. 2004) and 2MASS catalogues. Column 10
classifies the nebulae in the following conditional types: R --
round (e.g. Fig.\,\ref{fig:new}d and Fig.\,\ref{fig:new}g), E --
ellipsoidal (e.g. Fig.\,\ref{fig:new}a and Fig.\,\ref{fig:new}h), BP
-- bipolar (e.g. Fig.\,\ref{fig:new}l and
Fig.\,\ref{fig:J134233(24+8)}), BL -- bilateral (e.g.
Fig.\,\ref{fig:new}b and Fig.\,\ref{fig:new}k), C -- complex (e.g.
Fig.\,\ref{fig:J171116(24+5.8)} and
Fig.\,\ref{fig:J194353(24+3.6)}). Column 11 gives the angular size
of the nebulae. Comments on individual objects are given in column
12. This column also gives the IRAC $3.6 \, \mu$m magnitudes for
five CSs which are either marginally detected on 2MASS images or
visible only at the IRAC wavelengths (see
Section\,\ref{sec:search}); these magnitudes were extracted from the
GLIMPSE\,I Spring '07 Catalog and the GLIMPSE\,3D, 2007-2009 Catalog
\footnote{Both available at
http://irsa.ipac.caltech.edu/holdings/catalogs.html}.

\section{Spectroscopic confirmations and further work}
\label{sec:spectr}

\begin{figure*}
\begin{center}
\includegraphics[width=18cm,angle=0,clip=]{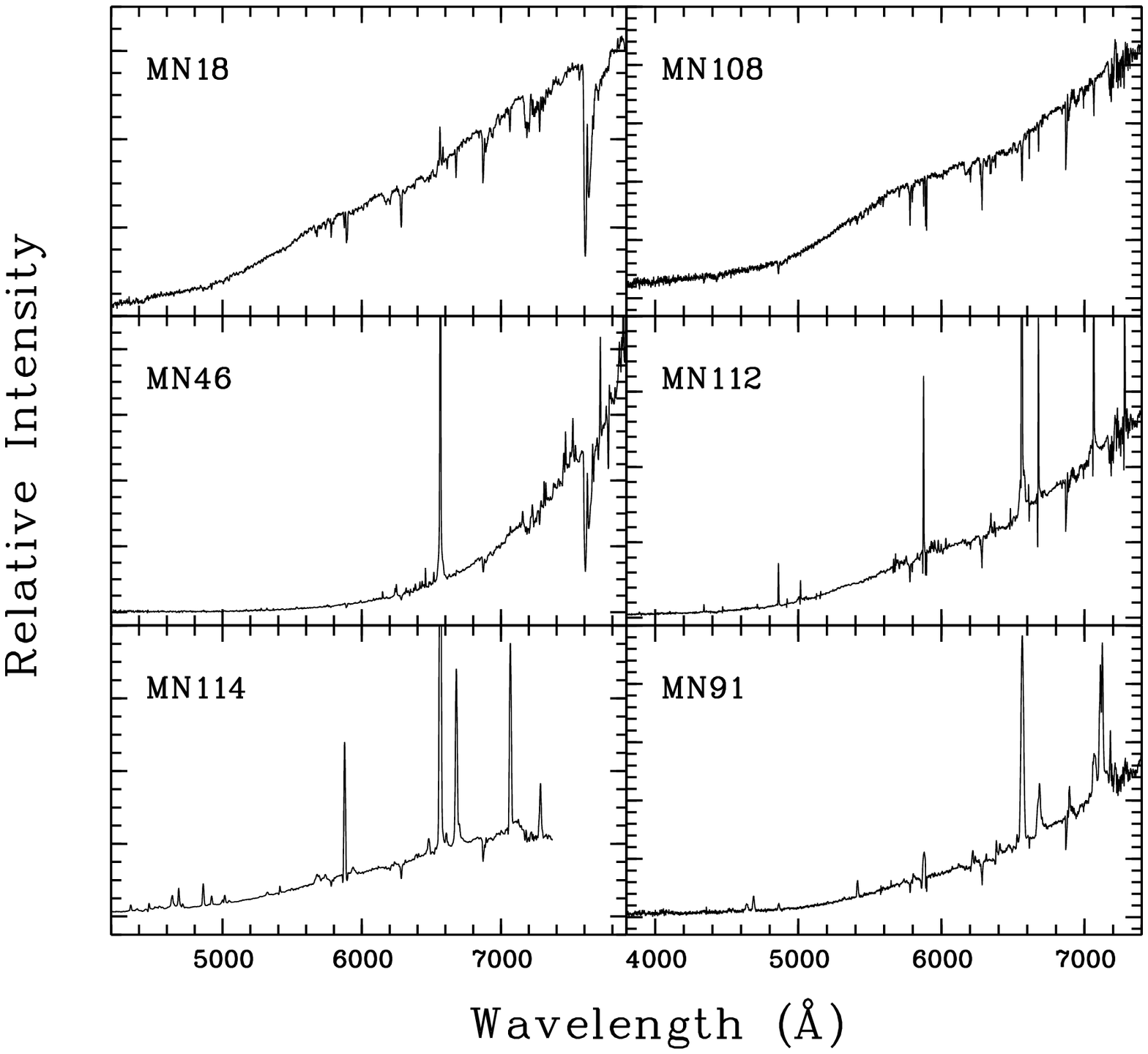}
\end{center}
\caption{Spectra of CSs of six nebulae discovered with {\it
Spitzer}. See text for details. } \label{fig:spectra}
\end{figure*}

To prove the availability of our method to search for evolved
massive stars via detection of their IR circumstellar nebulae
(Gvaramadze et al. 2009a,b,c), we took spectra of two dozens of CSs
associated with nebulae from our list (the details of these
observations will be presented in Gvaramadze et al., in preparation
and Kniazev et al., in preparation). Most of these stars turns out
to be either cLBVs, blue supergiants or WNL stars.

Fig.\,\ref{fig:spectra} shows a montage of six representative
spectra taken with the South African Astronomical Observatory 1.9-m
telescope (top and middle left panels), the Russian 6-m telescope
(bottom left panel) and the 3.5-m telescope in Calar Alto (right
panels). The obtained spectra were reduced using standard IRAF and
MIDAS routins and the nearby background spectra were subtracted. All
six stars are highly reddened objects with $A_V > 5-10$ mag so that
their spectra steeply rise towards the red and show numerous diffuse
interstellar bands. The high extinction towards the stars makes the
nebular contamination of their optical spectra negligible; in all
but one spectra the nebular emissions were not detected.

\begin{figure}
\begin{center}
\includegraphics[width=1.0\columnwidth,angle=0]{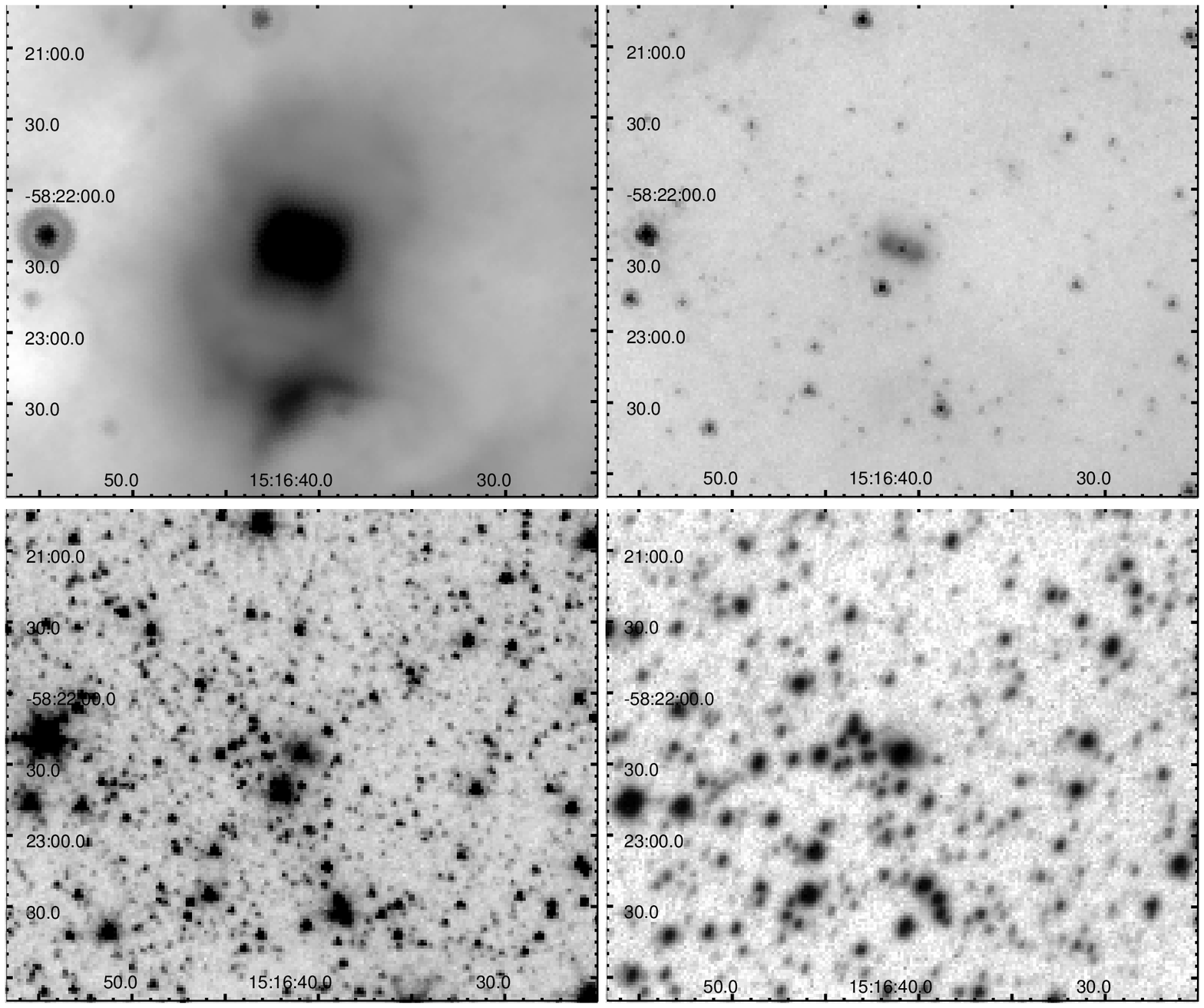}
\end{center}
\caption{MN18 at four wavelengths: {\it top left}: MIPS $24 \,
\mu$m; {\it top right} and {\it bottom left}: IRAC 8.0 and $3.6
\, \mu$m; {\it bottom right}: SHS.}
\label{fig:J151641(4band)}
\end{figure}

The first row gives spectra of two early OB supergiants associated,
respectively, with MN18 (Fig.\,\ref{fig:J151641(4band)}) and MN108
(Fig.\,\ref{fig:new}p). The spectrum of the CS of MN18 shows
numerous narrow absorption lines of H\,{\sc i} and He\,{\sc i}.
N\,{\sc ii} lines are also visible in absorption while the He\,{\sc
ii} absorptions are only marginally detected. We consider this star
as an early B supergiant. As we already mentioned in
Section\,\ref{sec:search}, MN18 is one of the two nebulae from our
list with detected optical counterparts (see
Fig.\,\ref{fig:J151641(4band)}). Correspondingly, the spectrum of
MN18 shows the prominent nebular emissions -- the H$\alpha$ line
flanked by the [N\,{\sc ii}] $\lambda\lambda \, 6548, 6584$ lines.
Note that MN18 was earlier classified as a PN by Kwok et al. (2008),
who named it as PNG 321.0-00.7. The spectrum of the CS of MN108
shows strong absorption lines of H\,{\sc i} and He\,{\sc i}.
Numerous He\,{\sc ii} absorption lines are also detected, of which
the strongest are $\lambda\lambda \, 4686$ and 5412. Most likely
this star is an O supergiant.

The second row gives spectra of two cLBVs [MN46
(Fig.\,\ref{fig:new}h) and MN112 (Fig.\,\ref{fig:new}q)]. The
spectrum of the CS of MN46 shows strong emission lines of hydrogen
with wide wings. The He\,{\sc i} emission lines are weak and wide.
The spectrum is remarkable by the presence of a forest of Fe\,{\sc
ii} and [Fe\,{\sc ii}] emission lines typical of LBVs during visual
minimum (e.g. Massey et al. 2007 and references therein). Overall,
the spectrum is quite reminiscent of that of the recently discovered
LBV star in the nuclear region of M33 (Valeev, Sholukhova \& Fabrika
2009). The spectrum of the CS of MN112 is almost identical to that
of the bona fide LBV P\,Cygni [see Gvaramadze et al. (2009c) for
details], while the nebula itself is reminiscent of that of the cLBV
GAL\,079.29+00.46 (see Section\,\ref{sec:search}). Like in P\,Cygni,
the spectrum is dominated by strong emission lines of H and He\,{\sc
i} and numerous permitted lines of N\,{\sc ii}, Fe\,{\sc iii} and
Si\,{\sc ii}. The only forbidden line detected in the spectrum is
the line of [N\,{\sc ii}] $\lambda \,5755$.

The third row gives spectra of two WNL stars associated with MN114
and MN91. These stars, WR\,138a and WR\,121b, were studied in detail
by Gvaramadze et al. (2009a,b). The spectrum of WR\,138a is
dominated by strong emission lines of H\,{\sc i}, He\,{\sc i} and
He\,{\sc ii}. Emission lines of N\,{\sc ii}, N\,{\sc iii}, C\,{\sc
iii}, N\,{\sc iv} and Si\,{\sc iv} are present as well. The
three-dimensional classification for WN stars by Smith et al. (1996)
suggests that WR\,138a belongs to the WN8-9h subtype (Gvaramadze et
al. 2009a). The spectrum of WR\,121b is similar in many respects to
that of WR\,138a. The effective temperature of this star, however,
is somewhat higher, which is manifested in weaker emissions of
H\,{\sc i}, much stronger emissions of N\,{\sc iv} $\lambda\lambda
\, 7111, 7123$ and the presence of the C\,{\sc iv} $\lambda \, 5808$
emission line. We identified WR\,121b as a WN7h star (Gvaramadze et
al. 2009b).

An additional prove of the availability of our method to search for
evolved massive stars comes from the fact that the CSs associated
with two nebulae from our list [namely MN23 (Fig.\,\ref{fig:new}d)
and MN43 (Fig.\,\ref{fig:new}g)] were identified as WR candidates on
the basis of their, respectively, near- and mid-IR colours (Mauerhan
et al. 2009) and emission in the near-IR narrow-band filters (Shara
et al. 2009), and confirmed as WN7 and WN5 stars in spectroscopic
follow-ups. We also searched for possible IR nebulae around other WN
stars discovered by these authors and by Hadfield et al. (2007) and
found that the WN9-10h star HDM1 (Hadfield et al. 2007) does show
signatures of a circumstellar nebula (most prominent at $8\, \mu$m).

\section{Conclusion and further work}

The results of preliminary spectroscopic study of CSs of compact
nebulae detected with {\it Spitzer} showed that the IR imaging
provides a powerful tool for revealing evolved massive stars and
suggested that a significant fraction of newly detected nebulae is
created by evolved massive stars. With our accepted proposals for
optical and infrared spectroscopy (with the Russian 6-m and the VLT
telescopes) of the remaining stars from our list we expect to
complete their spectral classification in the nearest future. We
expect also that the forthcoming spectrophotometric monitoring of
the already detected cLBVs accompanied by inspection of archival
plates will allow us to increase the known population of Galactic
LBVs. Needless to say that confirmation of the LBV nature of only
few objects from our list would have profound consequences for
understanding the LBV phenomenon and its role in the transition
between main-sequence O stars and hydrogen depleted WR stars (Conti
1976; Stahl 1986; Humphreys \& Davidson 1994; Langer et al. 1994;
Crowther, Hillier \& Smith 1995; Stothers \& Chin 1996; Pasquali et
al 1997; Smith \& Conti 2008).

\section{Acknowledgements}
We thank the Calar Alto Observatory for allocation of director's
discretionary time to this programme and the anonymous referee for
useful suggestions and comments allowing us to improve the content
and the presentation of the paper. VVG acknowledges the Deutsche
Forschungsgemeinschaft for partial financial support. AYK
acknowledges support from the National Research Foundation of
South Africa. The research has made use of the NASA/IPAC Infrared
Science Archive, which is operated by the Jet Propulsion
Laboratory, California Institute of Technology, under contract
with the National Aeronautics and Space Administration, the SIMBAD
database and the VizieR catalogue access tool, both operated at
CDS, Strasbourg, France.


\begin{thebibliography}{}
%
\bibitem{} Avedisova V.S., 1972, SvA, 15, 708
\bibitem{} Benjamin R.A. et al., 2003, PASP, 115, 953
\bibitem{} Billot N., Flagey N., Noriega-Crespo A., Shenoy S., Carey S., Mizuno D., Kraemer K., Latter B., 2009, BAAS, 41, 762
\bibitem{} Breysacher J., Azzopardi M., Testor G., 1999, A\&AS, 137, 117
\bibitem{} Brighenti F., D'Ercole, A., 1995, MNRAS, 277, 53
\bibitem{} Brighenti F., D'Ercole, A., 1997, MNRAS, 285, 387
\bibitem{} Carey S.J. et al., 2009, PASP, 121, 76
\bibitem{} Chen Y., Wang Z.-R., Qu Q.-Y., 1995, ApJ, 438, 950
\bibitem{} Chevalier R.A., Imamura J.N., 1983, ApJ, 270, 554
\bibitem{} Clark J. S., Larionov V.M., Arkharov A., 2005, A\&A, 435, 239
\bibitem{} Clark J.S., Egan M.P., Crowther P.A., Mizuno D.R., Larionov V.M., Arkharov A., 2003, A\&A, 412, 185
\bibitem{} Conti P.S., 1976, MSRSL, 9, 193
\bibitem{} Corral L.J., 1996, AJ, 112, 1450
\bibitem{} Crowther P.A., Hillier D.J., Smith L.J., 1995, A\&A, 293, 172
\bibitem{} Davies B., Figer D.F., Kudritzki R.-P., MacKenty J., Najarro F., Herrero A., 2007, ApJ, 671, 781
\bibitem{} Dolidze M.V., 1971, Astron. Tsirk., 629, 6
\bibitem{} Egan M.P., Clark J.S., Mizuno D.R., Carey S.J., Steele I.A., Price S.D., 2002, ApJ, 572, 288
\bibitem{} Esteban C., Smith L.J., Vilchez J.M., Clegg R.E.S., 1993, A\&A, 272, 299
\bibitem{} Fazio G.G. et al., 2004, ApJS, 154, 10
\bibitem{} Garcia-Segura G., Langer N., Mac Low M.-M., 1996b, A\&A, 305, 229
\bibitem{} Garcia-Segura G., Langer N., Mac Low M.-M., 1996a, A\&A, 316, 133
\bibitem{} Grosdidier Y., Moffat A.F.J., Blais-Ouellette S., Joncas G., Acker A., 2001, ApJ, 562, 753
\bibitem{} Gvaramadze V.V., Bomans D.J., 2008, A\&A, 490, 1071
\bibitem{} Gvaramadze V.V. et al., 2009a, MNRAS, 400, 524
\bibitem{} Gvaramadze V.V., Kniazev A.Y., Hamann W.-R., Berdnikov L.N., Fabrika S., Valeev A.F., 2009b, MNRAS,
in press (astro-ph/0911.0181)
\bibitem{} Gvaramadze V.V. et al., 2009c, MNRAS, in press (astro-ph/0912.5080)
\bibitem{} Johnson H.M., Hogg D.E., 1965, ApJ, 142, 1033
\bibitem{} Hadfield L.J., van Dyk S.D., Morris P.W., Smith J.D., Marston A.P., Peterson
D.E., 2007, MNRAS, 376, 248
\bibitem{} Helfand D.J., Becker R.H., White R.L., Fallon A., Tuttle S., 2006, AJ, 131, 2525
\bibitem{} Humphreys R.M., Davidson K., 1994, PASP, 106, 1025
\bibitem{} Homeier N.L., Blum R.D., Pasquali A., Conti P.S., Damineli A., 2003, A\&A, 408, 153
\bibitem{} Hora J.L., et al., 2009, BAAS, 41, 498
\bibitem{} Hutsemekers D., 1997, in Nota  A., Lamers  H.J.G.L.M., eds, ASP Conf. Ser. Vol. 120,
Luminous Blue Variables: Massive Stars in Transition. Astron. Soc.
Pac., San Francisco, p. 316
\bibitem{} King N.L., Walterbos R.A.M., Braun R., 1998, ApJ, 507, 210
\bibitem{} Kraemer K.E., et al., 2009, BAAS, 41, 467
\bibitem{} Kwok S., Zhang Y., Koning N., Huang H.-H., Churchwell E., 2008, ApJS, 174, 426
\bibitem{} Langer N., Hamann W.-R., Lennon M., Najarro F., Pauldrach A.W.A., Puls J., 1994, A\&A, 290, 819
\bibitem{} Lozinskaya T.A., Tutukov A.V., 1981, NInfo, 49, 21
\bibitem{} Marston, A.P., 1995, AJ, 109, 1839
\bibitem{} Marston A.P., Welzmiller J., Bransford M.A., Black J.H., Bergman P., 1999, ApJ, 518, 769
\bibitem{} Massey P., Waterhouse E., DeGioia-Eastwood K., 2000, AJ, 119, 2214
\bibitem{} Massey P., McNeill R.T., Olsen K.A.G., Hodge P.W., Blaha C., Jacoby G.H., Smith R.C., Strong
S.B., 2007, AJ, 134, 2474
\bibitem{} Mathis J.S., Cassinelli J.P., van der Hucht K.A., Prusti T., Wesselius P.R., Williams
P.M., 1992, ApJ, 384, 197
\bibitem{} Mauerhan J., Van Dyk S., Morris P., 2009, PASP, 121, 591
\bibitem{} McLean B.J., Greene G.R., Lattanzi M.G., Pirenne B., 2000, in Manset N., Veillet C., Crabtree D.,
eds, ASP Conf. Ser. Vol. 216, Astronomical Data Analysis Software
and Systems IX. Astron. Soc. Pac., San Francisco, p. 145
\bibitem{} Meynet G., Maeder A., 2003, A\&A, 404, 975
\bibitem{} Morris P., 2008, in Bresolin F., Crowther P.A., Puls J., eds, IAU Symp. 250, Massive Stars as Cosmic
Engines, p. 361
\bibitem{} Morris P.W., Stolovy S., Wachter S., Noriega-Crespo A., Pannuti T.G., Hoard
D.W., 2006, ApJ, 640, L179
\bibitem{} Nota A., Livio M., Clampin M., Schulte-Ladbeck R., 1995, ApJ, 448, 788
\bibitem{} Parker Q.A. et al., 2005, MNRAS, 362, 689
\bibitem{} Pasquali A., Langer N., Schmutz W., Leitherer C., Nota A., Hubeny I., Moffat
A.F.J., 1997, ApJ, 478, 340
\bibitem{} Phillips J.P., Ramos-Larios G., 2008b, MNRAS, 386, 995
\bibitem{} Phillips J.P., Ramos-Larios G., 2008c, MNRAS, 387, 407
\bibitem{} Phillips J.P., Ramos-Larios G., 2008a, MNRAS, 390, 1170
\bibitem{} Reipurth B., Schneider N., 2008, in Handbook of Star Forming Regions, Volume I: The Northern
Sky, ASP Monograph Publications, Vol. 4. Edited by Bo Reipurth, p.36
\bibitem{} Rieke G.H. et al., 2004, ApJS, 154, 25
\bibitem{} Robberto M., Ferrari A., Nota A., Paresce F., 1993, A\&A, 269, 330
\bibitem{} Shara M.M., Moffat A.F.J., Smith L.F., Niemela V.S., Potter M., Lamontagne R., 1999, AJ, 118, 390
\bibitem{} Shara M.M. et al., 2009, AJ, 138, 402
\bibitem{} Sholukhova O.N., Fabrika S.N., Vlasyuk V.V., 1999, AstL, 25, 14
\bibitem{} Sirianni M., Nota A., Pasquali A., Clampin M., 1998, A\&A, 335, 1029
\bibitem{} Skrutskie M.F. et al., 2006, AJ, 131, 1163
\bibitem{} Smith L.J., 1996, in Vreux J.M., Detal A., Fraipont-Caro D., Gosset E., Rauw G., eds, Wolf-Rayet
stars in the framework of stellar evolution. Liege: Universite de
Liege, Institut d'Astrophysique, p. 381
\bibitem{} Smith L.J., Crowther P.A., Prinja, R.K., 1994, A\&A, 281, 833
\bibitem{} Smith L.F., Shara M.M., Moffat A.F.J., 1996, MNRAS, 281, 163
\bibitem{} Smith N., 2007, AJ, 133, 1034
\bibitem{} Smith N., Conti P.S., 2008, ApJ, 679, 1467
\bibitem{} Smith N., Owocki S.P., 2006, ApJ, 645, L45
\bibitem{} Stahl O., 1986, A\&A, 164, 321
\bibitem{} Stothers R.B., Chin C.-W., 1996, ApJ, 468, 842
\bibitem{} Su\'{a}rez O., Garc\'{i}a-Lario P., Manchado A., Manteiga M., Ulla A., Pottasch S.R., 2006, A\&A, 458, 173
\bibitem{} Trams N.R., Voors R.H.M., Waters L.B.F.M., 1998, Ap\&SS, 255, 195
\bibitem{} Valeev A.F., Sholukhova O., Fabrika S., 2009, MNRAS, 396, L21
\bibitem{} van Genderen, A.M., 2001, A\&A, 366, 508
\bibitem{} van der Hucht K.A., 2001, New Astr. Rev., 45, 135
\bibitem{} van der Hucht K.A., 2006, A\&A, 458, 453
\bibitem{} Vanbeveren D., De Loore C., Van Rensbergen W., 1998, A\&AR, 9, 63
\bibitem{} Voors R.H.M. et al., 2000, A\&A, 356, 501
\bibitem{} Wachter S., Van Dyk S., Hoard D.W., Morris P., 2009, BAAS, 41, 721
\bibitem{} Weaver R., McCray R., Castor J., Shapiro P., Moore R., 1977, ApJ, 218, 377
\bibitem{} Weis K., 2001, RvMA, 14, 261
\bibitem{} Weis K., 2003, A\&A, 408, 205
\bibitem{} Werner M.W. et al., 2004, ApJS, 154, 1
\bibitem{} Zacharias N., Monet D.G., Levine S.E., Urban S.E., Gaume R., Wycoff G.L., 2004, AAS, 205, 4815

\end{thebibliography}
\end{document}